\documentstyle[12pt]{article}

\begin{document}
\thispagestyle{empty}
\begin{center}
\LARGE \tt \bf{Decay of inhomogeneities from the spin-torsion fluctuations in the early stages of inflation}
\end{center}
\vspace{1cm}
\begin{center} {\large L.C. Garcia de Andrade\footnote{Departamento de
F\'{\i}sica Te\'{o}rica - Instituto de F\'{\i}sica - UERJ
Rua S\~{a}o Fco. Xavier 524, Rio de Janeiro, RJ
Maracan\~{a}, CEP:20550-003 , Brasil.
e-mail.: garcia@dft.if.uerj.br}}
\end{center}
\vspace{1.0cm}
\begin{abstract}
The spin-torsion fluctuations are shown to act as source of decay of inhomogeneities in the early stages of inflationary de Sitter universe.This seems to be a resul of the repulsive nature of the spin-torsion effects.This result is shown through the investigation of the evolution equation of density perturbations in Einstein-Cartan cosmology.We also show that it is possible to place limits on torsion in the universe from the CMBR temperature anisotropy from the computations we made.The decay of inhomogeneities is not enough however to stop the structure formation as galaxies because the growth of inhomogeneitis dominates the torsion effects and the only effect of torsion seems to decrease very weakly the speed of formation of galaxies.
\end{abstract}
\vspace{1.0cm}       
\begin{center}
\end{center}
\newpage
\pagestyle{myheadings}
\paragraph*{}
The computation of cosmological density perturbations in General Relativity \cite{1,2} and in other theories of gravity \cite{3} have been used to investigate the gravitationally instability which leads to structure formation like galaxies and stars in the Universe.More recently the evolution equation of density primordial fluctuations have been undertaken by Palle \cite{4} and myself \cite{5}.In this note we apply these ideas to inflation to show that the evolution equation of density perturbations in the universe can be used to show that torsion effects are not important but in the early stages of inflationary \cite{6} de Sitter universe.This result have been also recently demonstrated by investigation of the methods of the inflaton potential reconstruction \cite{7}.The growth of inhomogeneities can be obtained in Einstein-Cartan (EC) cosmology the equation in two cases: first in the case of early inflation where Cartan torsion effects are important and  in the late stages where the torsion is redshift out.In general torsion effects are weak in the angular momentum of galaxies and other problems of structure formation and this seems to be the reason by which we should investigate torsion effects in the early and very early universe \cite{8}.Besides as is well known \cite{6} one of the merits of the inflation is to provide a theory of inhomogeneities in the universe through quantum fluctuations in the inflaton field about the vacuum fluctuations.Since the spin that appears in EC gravity is a quantum variable our investigation here seems to be fully justifiable.Inflation provides a redshift mechanism through inflation that acts on torsion in the same way that Linde \cite{9} showed to happen with monopoles and gravitational waves.The rpulsive nature of torsion in the early stages of inflation seems to appear in this kind of example.In fact we show that the dependence of the decay of inhomogeneity obtained here in the inflation process is the same sort of decay we obtain recently \cite{10} by analyzing the origin of the angular momentum of the galaxies in the context of EC gravity.Let us consider the Friedmann metric as 
\begin{equation}
ds^{2}=dt^{2}-a^{2}(t)({\delta}_{ik})dx^{i}dx^{k} 
\label{1}
\end{equation}
where ${i,k=1,2,3}$ and ${\delta}_{ik}$ is the Kronecker delta.In the case the background metric is de Sitter $a(t)=e^{H_{0}t}$ where $H_{0}$ is the Hubble constant and $t$ is the cosmic time.Besides $H_{0}=\frac{\dot{a}}{a}$.The EC cosmology equations are used to write as
\begin{equation}
H^{2}=\frac{8{\pi}G}{3}({\rho}-2{\pi}G{\sigma}^{2})
\label{2}
\end{equation}
and 
\begin{equation}
H^{2}+\dot{H}=-\frac{4{\pi}G}{3}({\rho}-8{\pi}G{\sigma}^{2})
\label{3}
\end{equation}
where the matter and spin are consider to be separatly conserved through the equations
\begin{equation}
{\dot{\rho}}=-3H{\rho}
\label{4}
\end{equation}
and
\begin{equation}
{\dot{{\sigma}^{2}}}=-6H{\sigma}^{2}
\label{5}
\end{equation}
where ${\sigma}^{2}=<S_{abc}S^{abc}>$ is the averaged squared spin density where $S_{abc}$ is the spin tensor and $(a=0,1,2,3)$.From the conservation equations and the Friedmann equation above we obtain the following expressions  
\begin{equation}
2H{\delta}H=\frac{4{\pi}G}{3}({\rho}{\delta}-2{\pi}G{\sigma}^{2})
\label{6}
\end{equation}
and
\begin{equation}
{\dot{\delta}}=-3{\delta}H
\label{7}
\end{equation}
The equation (\ref{5}) upon derivation becomes
\begin{equation}
2\dot{H}{\delta}H+2H{\dot{{\delta}H}}=-\frac{4{\pi}G}{3}(\dot{\rho}{\delta}-8{\pi}G\dot{{\sigma}^{2}})
\label{8}
\end{equation}
Here ${\delta}=\frac{{\delta}{\rho}}{\rho}$ is the contrast density and ${\delta}H$ represents the metric perturbation.Substitution of expression (\ref{7}) into (\ref{8}) yields
\begin{equation}
{\ddot{\delta}}+\frac{3}{2}({\rho}+\frac{2}{3}\dot{H})\dot{\delta}+{2{\pi}G}{\dot{\rho}}{\delta}=\frac{2{\pi}G}{3}{\dot{{\sigma}^{2}}}
\label{9}
\end{equation}
which is the equation of evolution of the density perturbation contrast, and shows clearly that it has the derivative of the spin-torsion density fluctuation ${\dot{{\delta}{\sigma}^{2}}}$ as the source of growing or decaying mode of inhomogeneities.To derive equation (\ref{9}) we have used also the approximation ${\delta}\dot{{\sigma}^{2}}={{\sigma}^{2}}-{{\sigma}^{2}_{0}}={{\sigma}^{2}}$ since the quantity ${\sigma}^{2}$ is is very small.This expression can be rewritten as
\begin{equation}
{\ddot{\delta}}+\frac{2}{3}(\frac{\dot{a}}{a}+\frac{2}{3}[\frac{\ddot{a}}{a}-(\frac{\dot{a}}{a})^{2}])\dot{\delta}-{6{\pi}G}M\frac{\dot{a}}{a^{4}}{\delta}=-2{\pi}^{2}G^{2}S^{2}_{0}\frac{\dot{a}}{a^{7}}
\label{10}
\end{equation}
where ${a,b=0,1,2,3}$.Now since we are performing the perturbations on a torsionless background we are able to use the general relativistic value for $a=t^{\frac{2}{3}}$ which yields 
\begin{equation}
{\ddot{\delta}}+\frac{2}{3}[\frac{2}{3t}+(\frac{4}{9t^{2}})]\dot{\delta}-\frac{{6{\pi}G}M}{t^{3}}{\delta}=-2\frac{{\pi}^{2}G^{2}S^{2}_{0}}{t^{5}}
\label{11}
\end{equation}
Let us now solve this equation is the particular case of the beginning of inflation.At this point we could imagine that the time tends to zero and then some terms can be dropped in this approximation leaving our equation as
\begin{equation}
{\ddot{\delta}}-\frac{6{\pi}GM}{t^{3}}{\delta}=-2\frac{{\pi}^{2}G^{2}S^{2}_{0}}{t^{5}}
\label{12}
\end{equation}
Since we can suppose that at the early inflation the growth of inhomogeneities due to torsion if existed would be extremely low we can drop the first term on the acceleration of perturbations and further  simplify equation  (\ref{12}) to
\begin{equation}
\frac{6{\pi}GM}{t^{3}}{\delta}=2\frac{{\pi}^{2}G^{2}S^{2}_{0}}{t^{5}}
\label{13}
\end{equation}
This equation now is easily solved to yield the solution
\begin{equation}
{\delta}=\frac{{\pi}G S^{2}_{0}t^{-2}}{3M}+c
\label{14}
\end{equation}
where c is a  constant of integration.This solution agrees with the solution found by Garcia de Andrade and A.Ribeiro \cite{10} for the density perturbation in the matter dominated era where we were trying to explain the origin of the angular momentum of galaxies by making use of a two fluid model in EC gravity.From this expression we also show  that is possible to place a limit on cosmological torsion from the COBE data.Since the COBE constraint places a limit for $\frac{{\delta}T}{T}<10^{-5}$ it is possible to obtain a limit from the torsion parameter $S^{2}_{0}$ from formula (\ref{14}).Since the inflation era starts at $t=10^{-35}s$, substituting this value into the equation (\ref{14})and obeying the COBE constraint one obtains a value of $10^{-17} cgs units$ for the spin-torsion parameter.De Sabbata and Sivaram \cite{11} have obtained the close value of $10^{-19}cgs units$ for this same quantity off the inflation era.Notice that this result is physically reasonable since on the inflation stage the rapid expansion of the universe makes the spin density to decrease and since the spin is connected algebraically with torsion in the EC gravity torsion also decrease which explains why our result is lower than de Sabbata and Sivaram result.To perform this computation we also consider $M=M_{Pl}=10^{33}g$ which is the Planck mass.Since time goes to infinity at the end of inflation we could drop other terms in the evolution equation to produce the equation
\begin{equation}
{\ddot{\delta}}+[\frac{4}{9t}+\frac{8}{27}t^{-2}]\dot{\delta}-6{\pi}GMt^{-3}{\delta}=0
\label{15}
\end{equation}
where torsion does not make any contribution.Dropping the second time derivative of ${\delta}$ as before 
\begin{equation}
[\frac{4}{9t}+\frac{8}{27}t^{-2}]\dot{\delta}-6{\pi}GMt^{-3}{\delta}=0
\label{16}
\end{equation}
which solution reads
\begin{equation}
{\delta}=t^{\frac{2}{3}-6{\pi}GM} 
\label{17}
\end{equation}
which is a growing mode for the inhomogeneities which appear in GR but that grows a little bit slower than the growing mode of torsionless case.The general solution to the evolution equation (\ref{11}) can be obtained with the help of the MAPLE V computer program and reads
\begin{equation}
{\delta}=9\frac{81A^{2}-414At-24A+644t^{2}}{t^{2}A(729A^{2}-648A+128)}+c_{1}t^{\frac{7}{9}}[Whittaker D(\frac{27A}{8})+\frac{280}{4374}t^{-1}]exp(\frac{4}{27t})+c_{2}[WhittakerW {\beta}{t}
\label{18}
\end{equation}
where $c_{1}$ and $c_{2}$ are integration constants and ${\beta}(t)$ is the same function of the first term on the RHS of equation (\ref{18}).From this expression we note that the spin-torsion term coincides with our previous results plus other decaying terms as ${\delta}_{spin-torsion}{\alpha}t^{-1}$ for example.Note also that these terms are all decaying terms and that growing terms that multiply Whittaker functions do not depend on spin-torsion effects.From the astronomical point of view this means that the presence of spin-torsion effects slow down the formation of galaxies instead of increasing them.This is due to the fact  that the torsion effects are certainly weaker than the curvature effects of GR.
\section*{Acknowledgments}
\paragraph*{}
I am very much indebt to Professors Ilya Shapiro,M.Gleiser and  R.Ramos for enlightening discussions on the subject of this paper.Financial supports from CNPq. (Brazilian Government Agency) and Fundacao de Amparo a pesquisa do Estado de Sao Paulo (FAPESP) are gratefully acknowledged. 

\end{document}